\documentclass[]{spie}  
\usepackage[]{graphicx}
\usepackage{amsmath,url, lscape}
\usepackage{amssymb}
\usepackage{aastex_hack}
\usepackage{wrapfig}

\title{Direct imaging of exoplanets in the habitable zone with adaptive optics} 

\author{Jared R. Males\supit{a,*}, Laird M. Close\supit{a}, Olivier Guyon\supit{a}, Katie Morzinski\supit{a,*},  Alfio Puglisi\supit{b}, Philip Hinz\supit{a}, Katherine B. Follette\supit{a}, John D. Monnier\supit{c}, Volker Tolls\supit{d}, Timothy J. Rodigas\supit{e}, Alycia Weinberger\supit{e}, Alan Boss\supit{e},  Derek Kopon\supit{f}, Ya-lin Wu\supit{a}, Simone Esposito\supit{b}, Armando Riccardi\supit{b}, Marco Xompero\supit{b}, Runa Briguglio\supit{b}, Enrico Pinna\supit{b}
\skiplinehalf
\supit{a}Steward Observatory, University of Arizona, Tucson, AZ, USA; 
\supit{b}Instituto Nazionale di Astrofisica, Osservatorio Astrofisico di Arcetri, Firenze, Italy;
\supit{c}Astronomy Department, University of Michigan;
\supit{d}Harvard-Smithsonian, CfA, Boston, MA, USA;
\supit{e}Carnegie Institution DTM, Washington, DC, USA;
\supit{f}MPiA Heidelberg, Germany;
\supit{*}NASA Sagan Fellow
}

\authorinfo{Further author information: (Send correspondence to J.R.M.)\\J.R.M.: E-mail: jrmales@as.arizona.edu\\Follow MagAO at \url{http://visao.as.arizona.edu/}}

\begin{document}

\maketitle

\begin{abstract}
One of the primary goals of exoplanet science is to find and characterize habitable planets, and direct imaging will play a key role in this effort.  Though imaging a true Earth analog is likely out of reach from the ground, the coming generation of giant telescopes will find and characterize many planets in and near the habitable zones (HZs) of nearby stars.  Radial velocity and transit searches indicate that such planets are common, but imaging them will require achieving extreme contrasts at very small angular separations, posing many challenges for adaptive optics (AO) system design.  Giant planets in the HZ may even be within reach with the latest generation of high-contrast imagers for a handful of very nearby stars.  Here we will review the definition of the HZ, and the characteristics of detectable planets there.  We then review some of the ways that direct imaging in the HZ will be different from the typical exoplanet imaging survey today.  Finally, we present preliminary results from our observations of the HZ of $\alpha$ Centauri A with the Magellan AO system's VisAO and Clio2 cameras.
\end{abstract}



\section{INTRODUCTION}
\label{sec:intro}

Direct imaging of exoplanets in the habitable zone (HZ) -- the region where liquid water could exist on the surface of a planet -- is typically assumed to be in the realm of space missions.  When it comes to Earth-sized planets orbiting Sun-like stars this is a correct assumption.  The $\sim$$10^{-10}$ contrast required to image such a planet is likely impossible to achieve through the ravages of atmospheric turbulence.  This is not the end of the story, however. Not all stars are Sun-like, and not all planets are Earth-sized.  Ground-based imaging using Adaptive Optics (AO) will play a role in our explorations of the HZ.

As of this writing there are 1810 confirmed exoplanets\footnote{2014/07/11, \url{http://exoplanets.eu/}}, most found by either the radial velocity (RV) or transit techniques.  In addition, the \emph{Kepler} mission has another 4234 planet candidates waiting to be confirmed\footnote{\url{http://www.nasa.gov/mission_pages/kepler/main/}}, most of which are likely true planets\cite{2013ApJ...766...81F}.  Moreover, the \emph{Kepler} results show that the average number of planets per star is \emph{at least} one\cite{2013ApJ...766...81F}.  Most important to this discussion is that more than $50$ of these planets are currently believed to orbit their stars within the HZ\cite{2012PASP..124..323K}.  However, most of these are extrasolar gas giants (EGPs), with radii and masses much larger than Earth.

Recently, the \emph{Kepler} team announced the discovery of the first Earth-radius planet orbiting within the HZ of a main sequence star, Kepler-186f\cite{2014Sci...344..277Q}.  This planet's host star is an M1 dwarf, far from being 
Sun-like.  This result is not surprising: the occurrence rate of planets is higher around cooler stars\cite{2012ApJS..201...15H}, with short period planets being roughly twice as common around M stars as around G stars\cite{2014arXiv1406.7356M}.  In fact, it has been estimated that the nearest habitable planet is within 5 pc of the sun\cite{2013ApJ...767...95D}, but most likely orbiting an M dwarf.

Guyon et al (2012)\cite{2012SPIE.8447E..1XG} analyzed a catalog of nearby stars, finding 300 stars around which, given their luminosities and distances, future 30 m class extremely large telescopes (ELTs) will be able to characterize potentially habitable super-Earths, mainly around cooler stars.  Similarly, Crossfield (2013)\cite{2013A&A...551A..99C} used statistics from the \emph{Kepler} mission to model the population of planets possibly orbiting stars within 8pc, and estimated that ELTs will image approximately 10 short period planets around those stars.  These studies show that ground-based imaging in the HZ will be an important part of future exoplanet studies.

In what follows we will not actually discuss AO or coronagraphy or any of the technological details required to image planets in the HZ.  Rather, we hope to make a convincing case that given the current and future capabilities of AO systems, direct imaging in the HZ from the ground will be possible.   Information about such capabilities is available in these proceedings and elsewhere.  Current high contrast imaging AO systems and instruments include: the Gemini Planet Imager (GPI)\cite{Macintosh12052014, 2014SPIE_macintosh} instrument; the Spectro-Polarimetric High-contrast Exoplanet REsearch (SPHERE)\cite{2014SPIE_beuzit} instrument; Project 1640\cite{2014SPIE_vasisht}; the Subaru Coronagraphic Extreme AO (SCExAO)\cite{2014SPIE_jovanovic} system; the Large Binocular Telescope (LBT) Interferometer (LBTI)\cite{2014SPIE_hinz} instrument; and the Magellan AO (MagAO)\cite{2014SPIE_morzinski} system.  The nature and names of future projects are constantly changing, but some references with detailed predictions about future capabilities include: the Planet Formation Imager (PFI)\cite{2006SPIE.6272E..20M} for the Thirty Meter Telescope (TMT); the TIGER\cite{2012SPIE.8446E..1PH} instrument for the Giant Magellan Telescope (GMT); and the Planetary Camera and Spectrograph (PCS)\cite{ao4elt3_12804} for the European ELT.

We first briefly review the definition of the HZ, and begin discussing some of the ramifications of the HZ for direct imaging.  We then analyze the expected brightness of exoplanets in reflected light in the HZ, and show that both large planets around solar type stars and small planets around cooler stars are detectable from the ground.  We then turn to a detailed discussion of the detectability of EGPs, and argue that for some very nearby stars EGPs are detectable using the latest generation of AO systems.  Next, we discuss some of the challenges that the relatively rapid orbital motion of HZ planets present for direct imaging.  Finally, we briefly introduce the initial survey of the HZ of the nearest Sun-like star $\alpha$ Cen A using MagAO in both the visible and thermal infrared (IR).

\section{THE HABITABLE ZONE}
\label{sec:hz}
Here we will briefly review the concept of the HZ, and describe how it is defined.  For a more thorough introduction to the HZ we refer the reader to recent reviews by Seager (2013)\cite{2013Sci...340..577S} and Kasting et al (2013)\cite{Kasting25112013}.  The HZ is, by convention, defined as the region around the star where the surface temperature of a terrestrial planet allows the presence of liquid water\cite{2013Sci...340..577S}.  Liquid water is a requirement for all life on Earth, and has many physical and chemical properties which make it an ideal medium for life.  For our purposes, the locations of the inner and outer boundaries of the liquid water HZ are fundamental inputs into the requirements that must be met by an AO system in order to image planets there.  

\subsection{The Earth-Like Habitable Zone}
We first consider definitions of the HZ which focus on planets very similar to Earth, in terms of size, composition, and atmosphere.  An empirical estimate for the limits of the HZ was proposed by Kasting et al (1993)\cite{1993Icar..101..108K}.  Considering the last time Venus and Mars appear to have had liquid water on their surfaces, and accounting for the evolution of solar luminosity with time, these limits are:
\vspace{-3mm}
\begin{itemize}
\item Inner Edge (recent-Venus): 0.75 AU
\item Outer Edge (early-Mars): 1.77 AU
\end{itemize}
\vspace{-3mm}
More detailed modeling of Earth's climate and geological processes leads to a more restrictive HZ.  A generally accepted inner edge is set by photodissociation of water with subsequent hydrogen escape.  A generally accepted outer edge is set by the maximum distance where the CO$_2$ greenhouse effect can maintain surface temperatures.  Recent estimates for these limits from Kopparapu et al (2013)\cite{2013ApJ...765..131K} are
\vspace{-3mm}
\begin{itemize}
\item Inner edge (H$_2$O loss): 0.99 AU
\item Outer edge (Max. greenhouse): 1.70 AU
\end{itemize}
\vspace{-3mm}
Note that there is great deal of ongoing work on this topic, and these values will likely be revised under further study.  Finally, we note the ``wide HZ'' given by Traub (2011)\cite{2012ApJ...745...20T}.  These limits, attributed to Kasting,  are:
\vspace{-3mm}
\begin{itemize}
\item Inner edge: 0.72 AU
\item Outer edge: 2.00 AU
\end{itemize}
\vspace{-3mm}
This ``wide-HZ'' is motivated by ensuring that no potentially habitable planets are missed in data analysis and interpretation.   When considering these different definitions of the HZ and the requirements they place on future AO systems, the most challenging limits should be used when designing an instrument for direct imaging in the HZ.  This may not simply be the closest inner edge --- closer planets have more favorable contrasts, for instance.  

\subsection{The Wider Habitable Zone}
Many studies have relaxed the requirement of being specifically Earth-like, and proposed much wider HZs.  For instance Zsom et al (2013)\cite{2013ApJ...778..109Z} considered very low humidity atmospheres, and found that such a planet could retain liquid water as close as 0.5 AU (though see discussion in Kasting et al (2013)\cite{Kasting25112013}).  Planet mass also likely plays a role.  More massive planets  are expected to have denser atmospheres, leading to a weaker greenhouse effect, and hence the inner edge of the HZ will be closer\cite{2014ApJ...787L..29K}.  The reverse holds for lower mass planets.

Turning to even less Earth-like environments, Pierrehumbert and Gaidos (2011)\cite{2011ApJ...734L..13P} argue that molecular hydrogen is a very effective greenhouse gas due to collision-induced absorption.  For a 3 $M_\earth$ planet with a 40 bar H$_2$-He atmosphere at 10 AU from a G star, their calculations show that surface temperatures would be $\sim$$280$ K.  With these much wider separations such planets may be important targets of future direct imaging efforts.

A dense H$_2$ atmosphere may even be able to keep the surface of a free floating planet warm enough to harbor liquid water\cite{1999Natur.400...32S}.  Another source of energy which could keep water liquid far from the star is tidal heating of a moon, which is evident in our solar system on bodies such as Europa\cite{2000Sci...289.1340K} and Enceladus\cite{Iess04042014}.  With the various ways it is possible to keep temperatures high enough for liquid water, one can argue that the outer edge of the HZ extends to infinity\cite{2013ApJ...778..109Z}.

\subsection{Other HZ Concepts}
Before moving to a discussion of direct imaging in the HZ, we note a few more important HZ concepts.  Main sequence stars become more luminous as they evolve, which means the location of the HZ must also evolve, possibly limiting the development of life.  The continuously habitable zone  (CHZ) concept applies the above definitions with the additional restriction that the planet can support liquid water for a significant period of time, even when faced with stellar evolution\cite{1978Icar...33...23H,1993Icar..101..108K}.    We are also generally neglecting any discussion of biosignatures, that is the signatures of life that we might observe.  We refer to the review articles cited above, as well as the chapter on terrestrial planet atmospheres in \emph{Exoplanets} by Meadows and Seager (2011)\cite{2011exop.book..441M}, for detailed discussions of these topics.

\subsection{The Direct Imager's Habitable Zone}

The above discussion leads to a few points of interest when considering direct imaging in and near the HZ.  First, for the remainder of this article we will assume that the location of the HZ is loosely set by
\begin{equation}
a_{HZ} \approx \sqrt{\frac{L_*}{L_\sun}} \mbox{ AU}.
\label{eq:ahz}
\end{equation}
so that we can apply the HZ concepts to different spectral types.  We have here defined the 1 AU equivalent distance from the star, but we can also scale the inner and outer edges of the HZ by $a\propto\sqrt{L_*}$.  This should not be done lightly, though, as it ignores the impact of the stellar spectrum on the planet's climate and implicitly assumes that the Bond albedo is constant\cite{Kasting25112013}.

In the coming sections we will discuss some of the impacts that moving into the HZ has on direct imaging of exoplanets.  Many of these affect both space-based and ground-based efforts.  The most important characteristic of direct imaging in the HZ is that irradiation by the star is important (essentially by definition).  In this regime, light from the star that is reflected from the planet will play a major role in the brightness of the planet.  In addition, the incoming radiation from the star heats the planet so that the planet's age and mass do not control its luminosity in the thermal infrared.  

Planets in the HZ are much closer to their stars than the planets currently targeted through direct imaging.  This means that orbit dynamics will play a more significant role in the HZ.  In reflected light, the brightness of the planet depends on orbital phase, such that a planet may only be detectable for a fraction of its orbit.  Direct imaging instruments have a finite inner working angle (IWA), which means that planets too close to their star will not be detectable.  Orbital time scales will be short enough in the HZ, particularly around low-mass stars, that whether or not a planet is detectable will change rapidly for both of these reasons.  On large telescopes, orbital motion projected onto the focal plane can be fast enough that the resultant smearing will limit our sensitivity. 

Finally, we note that not all planets in the HZ will be habitable.  Ice and gas giant planets are generally not considered to be habitable themselves -- even if in the HZ -- but they may still be astrobiologically interesting if they have moons\cite{2013AsBio..13...18H}.   In the context of direct imaging from the ground with AO, such planets are interesting because (as we will show below) their larger radii make them detectable.  Precursor searches from the ground with AO will be able to identify such planets, allowing space telescopes to avoid those stars and thus concentrate precious telescope time on stars which may still have a terrestrial planet in the HZ.  

\section{CONTRAST IN REFLECTED LIGHT}
\label{sec:contrast_refl}
In the HZ, starlight reflected from the planet will play a major role in direct imaging detections.  Especially for smaller radius planets, it may be the only way in which they are detected.  The ratio of reflected flux from the planet to the flux from the star, the contrast, is given by\cite{2005ApJ...627..520S,2010ApJ...724..189C}  
\begin{equation}
\frac{F_p}{F_*} = 1.818\times10^{-9} \left(\frac{R_p}{1R_{\earth}}\times \frac{1\mbox{ AU}}{a}\right)^2 A_g \Phi(\alpha)
\label{eq:fpfstar}
\end{equation}
where $F_p$ and $F_*$ are flux from the planet and star respectively, $R_p$ is planet radius, $a$ is the separation of the planet from its star, and $A_g$ is the planet's geometric albedo.  The orbital phase angle $\alpha$ at the time of observation  is given by\cite{2005ApJ...627..520S}
\begin{equation}
\cos(\alpha) = \sin(f + \omega)\sin(i)
\end{equation}
where $f$ is the true anomaly, $\omega$ is the argument of periastron, and $i$ is the inclination of the orbit.  Finally $\Phi(\alpha)$ is the classical phase function which describes how much light, relative to full phase ($\alpha = 0$), is reflected by the planet.  As an example, if we assume isotropic pure scattering we have the Lambert phase function\cite{2011exop.book..419B}
\begin{equation}
\Phi(\alpha) = \frac{\sin(\alpha) + (\pi - \alpha)\cos(\alpha)}{\pi}.
\label{eq:lambert_phase}
\end{equation}
Real atmospheres are more complicated, and so is the resultant $\Phi(\alpha)$ \cite{2005ApJ...627..520S,2010ApJ...724..189C}.

\subsection{Contrast of an Earth}
As a starting point, we assume a full-phase albedo $A_g=0.3$ and use the Lambert phase function for $\alpha = 90^o$, which gives $\Phi(\alpha) = 0.32$.  For an Earth-like planet, i.e. $R_P = 1 R_\earth$, at 1 AU this yields a contrast of $1.7\times10^{-10}$.  It is almost certain that such a planet can not be detected from the ground. Traub and Oppenheimer (2011)\cite{2011exop.book..111T} present an analysis which shows that given reasonable assumptions about atmospheric turbulence, photon noise will prevent us from controlling the wavefront with sufficient accuracy to achieve this contrast.  Cavarroc et al (2006)\cite{2006A&A...447..397C} demonstrated that even with a 100 hr exposure on a future 100 m telescope, this contrast will not be achieved at tight angular separations because of residual non-common path aberrations.  A space telescope is almost certainly required to detect light reflected from an Earth-like planet orbiting a Sun-like star.

The situation is not hopeless, however.  The dependence of contrast on $R_p^2$, and the relationship between contrast, separation, and the location of the HZ for different spectral types means that ground-based AO will still have a role to play in the HZ.

\subsection{Detectability in the HZ}
\label{sec:coolstars}

Combining the scaled location of the HZ, Equation (\ref{eq:ahz}), with the reflected light contrast, Equation (\ref{eq:fpfstar}) we find that HZ contrast is dependent on planet radius and stellar luminosity as
\begin{equation}
\frac{F_p}{F_*}\left(a_{HZ}\right) \propto \frac{R_p^2}{L_*}.
\label{eq:fpfstar_hz}
\end{equation}

\begin{figure}
\begin{center}
\includegraphics[height=7cm]{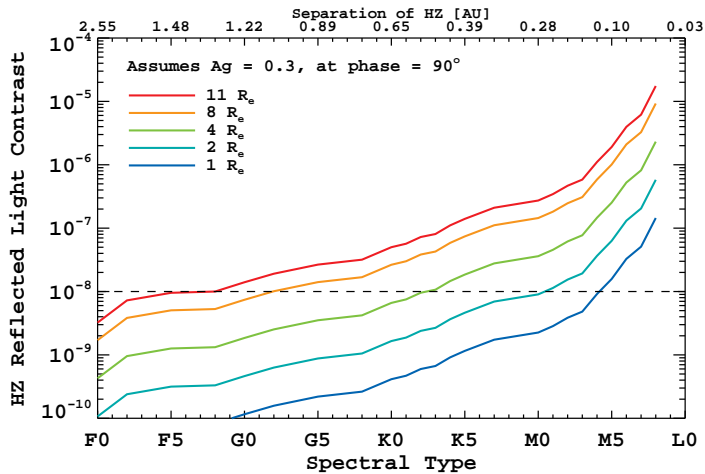}
\end{center}
\caption{The reflected light contrast of planets in the 1 AU equivalent HZs of main sequence stars, showing the dependence of relative planet brightness on both spectral type and planet size.  Cooler, less luminous, stars have closer HZs, which results in lower contrasts.  Note that the top axis shows separation of the HZ, using Equation (\ref{eq:ahz}).  The dashed line at $10^{-8}$ contrast indicates the ground-based contrast limit we adopt for extreme AO systems.  
\label{fig:ms_hz_loc}}
\end{figure}

Luminosity is generally a well defined function of main sequence spectral type.  Here we use the data tabulated in Lang (1992)\cite{1992adps.book.....L}, though other references can vary by $\pm50\%$ or more, especially for early spectral types.  Given the exponential nature of the luminosity sequence, and of contrast itself, this of minor concern here.  In Figure \ref{fig:ms_hz_loc} we show the evolution of contrast in the 1 AU equivalent HZ versus main sequence spectral type.  For this analysis we have again assumed $A_g=0.3$, $\alpha = 90^o$, and used Equation (\ref{eq:lambert_phase}).

Following Guyon et al. (2012)\cite{2012SPIE.8447E..1XG}, we adopt $10^{-8}$ as the achievable contrast limit for extreme AO systems.  To be sure this is a challenging contrast level for ground-based observations, but is typical of predictions made for future ELT class extreme AO systems\cite{2006SPIE.6272E..20M}.  For a similar but more detailed analysis of contrast limits see Crossfield (2013)\cite{2013A&A...551A..99C}.  As the Figure shows, ground based ELTs will be able to detect giant planets around Sun-like stars, and will detect Earth-sized planets around cooler M dwarfs.  Also note that ELT-like resolutions will be required due to the comparatively small separations of the HZ of M dwarfs.

Using Figure \ref{fig:ms_hz_loc} we can determine when a HZ planet of a given size becomes detectable from the ground as a function of spectral type.  Two effects are at play here.  The first is that the HZ of less luminous stars is at a closer separation, and so planets located there have lower contrast (i.e. are easier to detect).  The second is that larger planets are of course brighter for a given $A_g$.  In Table \ref{tab:hzcontrasts} we give the spectral type at which planets of various sizes become detectable in the HZ at $10^{-8}$ contrast under the albedo and phase assumptions we have made.  Note that the uncertainty in these determinations is roughly $\pm1$ spectral type, driven mainly by scatter in main sequence luminosities.

\begin{table}
\begin{center}
\begin{tabular}{lccc}\hline
Planet  Radius               &     Sp. Type    &    Luminosity  &   Separation\\
 \hspace{5mm} $[R_\earth]$   &                 &    [ $L_\sun$ ]  &     [AU]    \\ 
\hline
\hline
1 (Earth)                    &     M4          &      0.019     &    0.13     \\
2 (Super-Earth)              &     M0          &      0.077     &    0.27     \\
4 (Neptune)                  &     K2          &      0.29      &    0.53     \\
6                            &     G8          &      0.66      &    0.80     \\
8 ($\sim$0.05 $M_{Jup}$ EGP) &     G2          &       1.1      &    1.06     \\
9 (Saturn)                   &     G0          &       1.5      &    1.19     \\
11 (Jupiter)                 &     F8          &       2.1      &    1.45     \\
\hline
\end{tabular}
\end{center}
\caption{Habitable zone detectability at $10^{-8}$ contrast.
\label{tab:hzcontrasts}
}
\end{table}

\section{EXTRASOLAR GIANT PLANETS IN THE HZ}
\label{sec:egps}

An important physical property of EGPs and brown dwarfs (BDs) is that the radius of such an object only weakly depends on its mass.  The structure of planets with $M \lesssim 3 M_{Jup}$ is supported by Coulomb pressure, and for planets with $M \gtrsim 3 M_{Jup}$ the support is provided by degeneracy.  The result is that all EGPs and BDs are $1 R_{Jup}$ in radius to within $\sim\pm15\%$\cite{2007ApJ...659.1661F,2011ApJ...736...47B}.  For EGPs with $M \lesssim 1M_{Jup}$ a useful scaling between mass and radius is\cite{2011exop.book..397F}
\[
\frac{R}{R_{Jup}} = \left( \frac{M}{M_{Jup}} \right)^{1/10}.
\]
Observational evidence is mounting that this relationship can hold to very low masses.  Batygin and Stevenson (2013)\cite{2013ApJ...769L...9B} considered the case of Kepler-30d, an $M=23.1\pm2.7M_\earth$, $R=8.8\pm0.5R_\earth$ transiting planet discovered by the \emph{Kepler} satellite.  Though Kepler-30d's mass  is close to that of Neptune, its radius is nearly that of Saturn.  In fact, it is nearly perfectly described by the $R=M^{1/10}$ scaling.

\begin{wrapfigure}[20]{r}{4in}
\vspace{-0.5in}
\begin{center}
\includegraphics[width=7cm,angle=90]{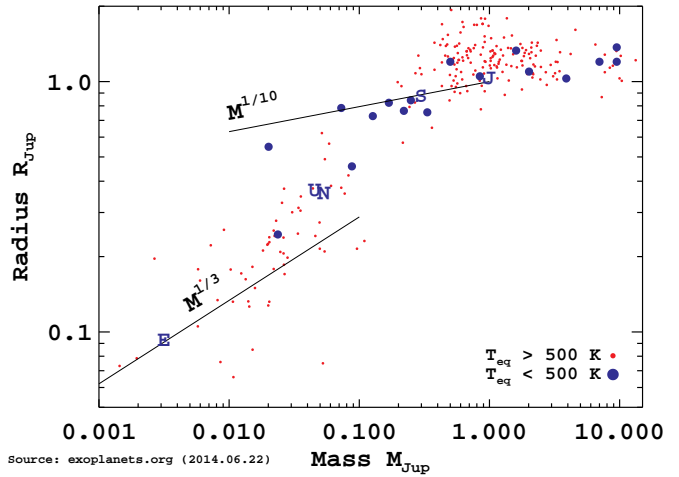}
\end{center}
\vspace{-.2in}
\caption{Mass-radius diagram for exoplanets which have had both mass and radius measured.  Data from exoplanets.org.  The large blue dots denote planets with $T_{eq} < 500$ K, assuming $A_B = 0$.  Though not a rigorously derived cutoff, this shows that the planet radii here are not simply an artifact of extreme irradiation.   
\label{fig:egp_mr}}
\end{wrapfigure}

A growing number of planets have had both radius and mass measured by the transit and RV techniques, respectively.  Since the transit gives a precise measure of inclination, the $\sin(i)$ uncertainty in mass from RV is eliminated.  In Figure \ref{fig:egp_mr}, we show a mass-radius diagram using data extracted from the exoplanets.org\cite{2011PASP..123..412W} database.  Using values for $R_*$ and $T_*$ provided by the database we calculated stellar luminosity, and from there estimated equilibrium temperature of the planet according to
\begin{equation}
\label{eq:teq}
T_{eq} = \left[\frac{1-A_B}{f}\right]^{1/4} 278.5\mbox{K} \left(\frac{L}{L_{\sun}}\right)^{1/4}\left(\frac{a}{1\mbox{AU}}\right)^{-1/2}
\end{equation}

\noindent where we have (for now) assumed that the Bond albedo is $A_B=0$ and the distribution factor is $f=1$.  In Figure \ref{fig:egp_mr} we also show the $M^{1/10}$ line scaled from Jupiter, as well as the $M^{1/3}$ line --- appropriate for rocky planets --- scaled from Earth.  Planets with masses at least as low as $\sim$$0.05 M_{Jup}$ ($\sim$$15 M_\earth$) appear to have radii of $\sim$$8 R_{\earth}$, placing them on the $M^{1/10}$ relation.

\begin{figure}
\begin{center}
\includegraphics[width=8cm]{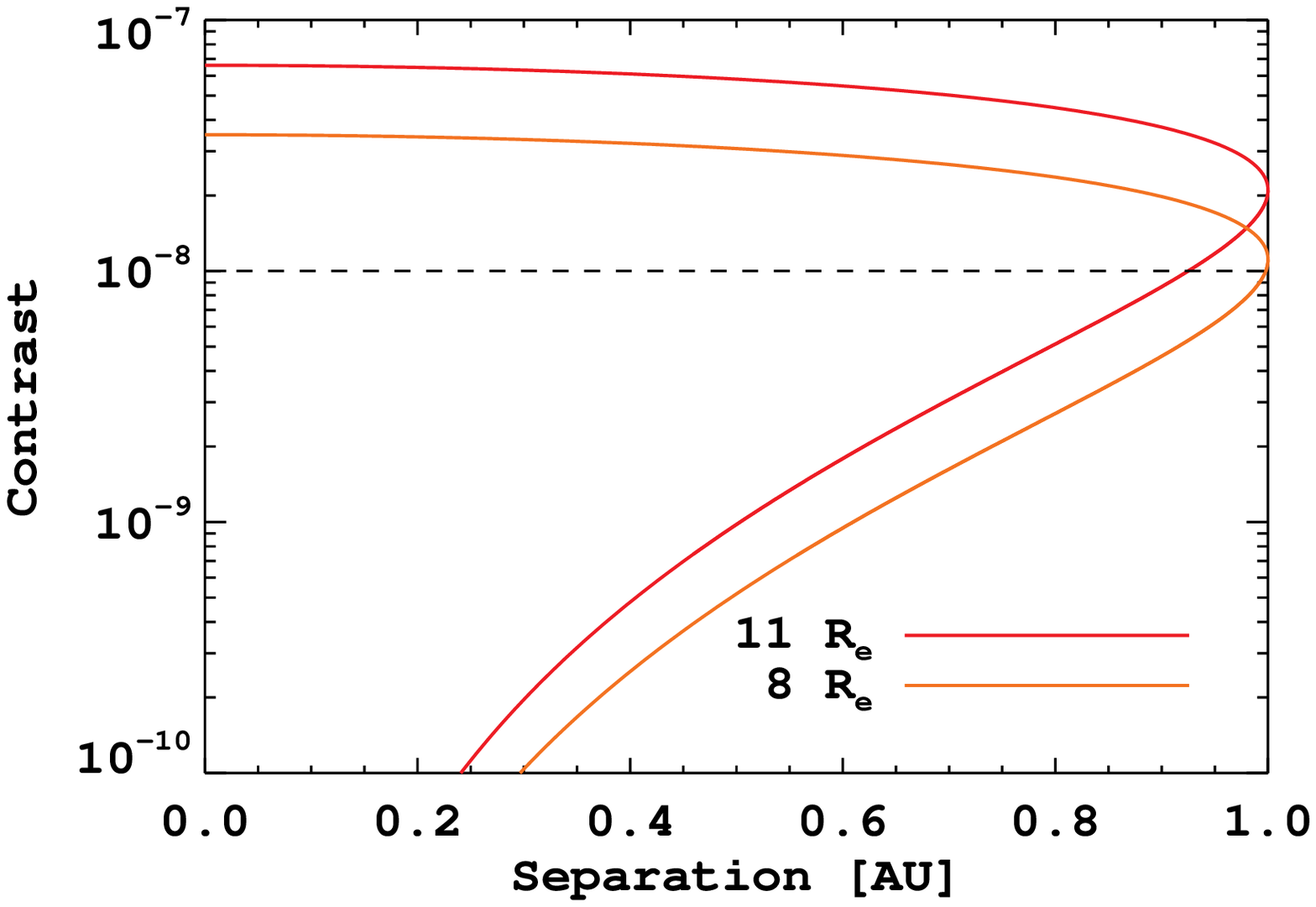}
\includegraphics[height=8cm,angle=90]{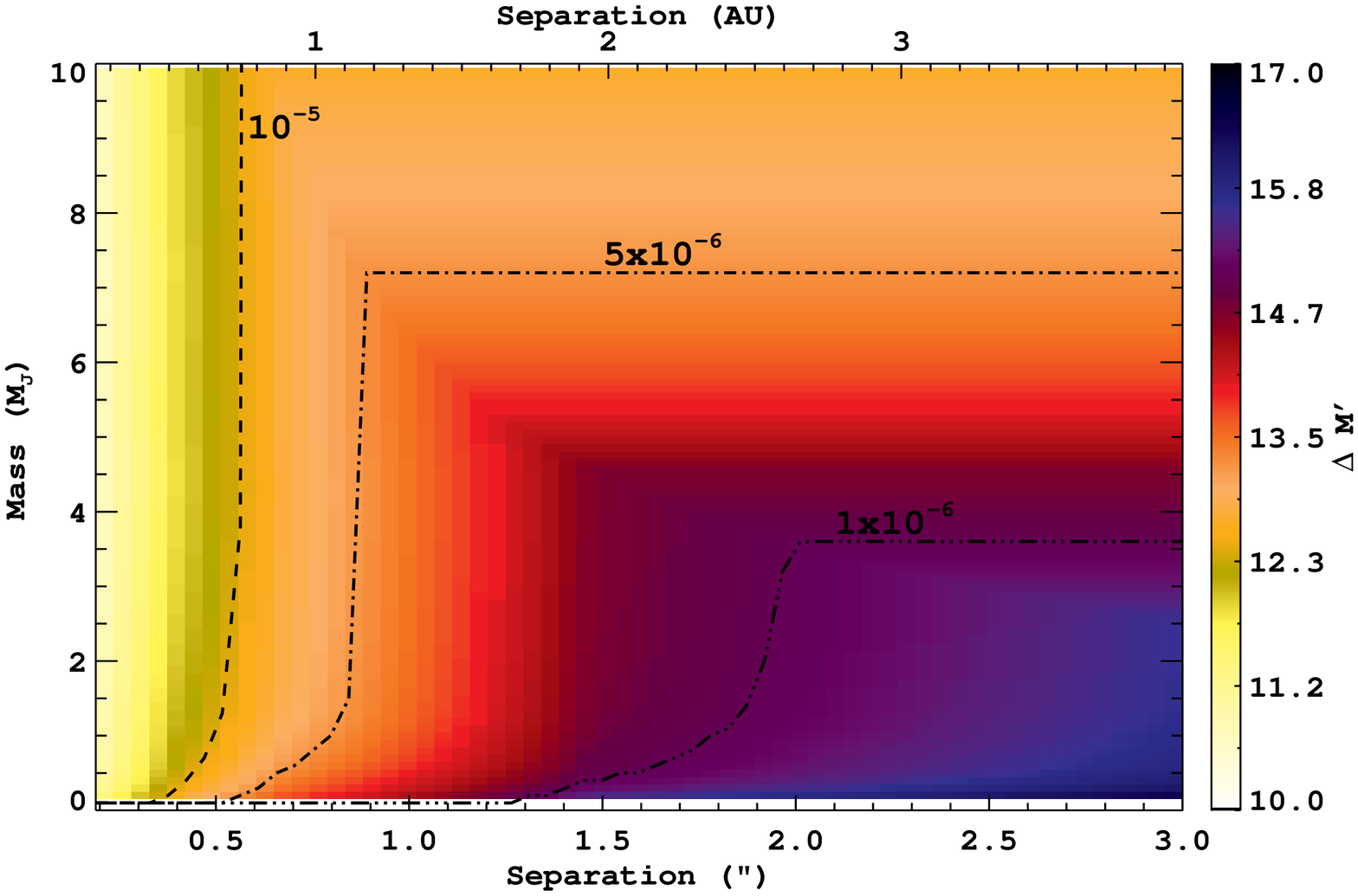}
\end{center}
\caption{Left: reflected light contrasts of EGPs with radii of $11 R_{\earth}$ and $8 R_{\earth}$.  The curves show contrast vs. projected separation over the course of an orbit, where orbital phase is changing. The weak dependence of radius on mass means that such planets span nearly 2 orders of magnitude in mass, but only a factor of 2 in reflected light contrast (see Equation (\ref{eq:egp_cont_mass})).  Right: a COND model based prediction for the $M'$ ($4.7 \mu$m) contrast of EGPs orbiting $\alpha$ Cen A.  The required contrasts are achievable using even today's AO technology.  Together these plots show that in and near the HZ, even very low mass (10 to 20 $M_\earth$) EGPs will be readily detectable from the ground with AO.
\label{fig:egp_detect}}
\end{figure}

These results have important implications for the detectability of EGPs in the HZ.  First we note that reflected light contrast of EGPs goes as
\begin{equation}
\frac{F_p}{F_*} \propto R_p^2 \propto M_p^{2/10}.
\label{eq:egp_cont_mass}
\end{equation}
In Figure \ref{fig:egp_detect} we show projected orbital separation vs. contrast for EGPs with radius $11R_{\earth}$ ($\sim$$1R_{Jup}$) and $8R_{\earth}$.  We assumed the same $A_g$ and $\alpha$ as above, and placed the planet 1 AU from a G2 star.  Such planets span nearly 2 orders of magnitude in mass, but the result is only a factor of 2 in contrast in the HZ.  

A similar argument holds for EGP brightness in the thermal infrared.  We expect luminosity to vary as
\begin{equation}
L_p \propto R_p^2 T_p^4.
\end{equation}
In and near the HZ the minimum planet temperature $T_p$ will be controlled by irradiation from the star, rather than mass and age.  Since $R_p$ depends only weakly on mass, luminosity will be only weakly dependent on mass.  Figure \ref{fig:egp_detect}b shows this.  Here we have used the COND models\cite{2003A&A...402..701B} as follows: first, the temperature $T_{evol}$ and radius of a planet at a given mass are determined from the associated evolutionary sequence for the star's age.  Then, at a given separation, $T_{eq}$ was determined using Equation (\ref{eq:teq}) with $A_B=0.3$ and $f=1$.  Now for each mass and separation pair, if $T_{evol} > T_{eq}$ then the planet's evolutionary temperature was used.  If instead $T_{eq} > T_{evol}$ then the equilibrium temperature was used.  This analysis was extended to lower masses using the $M^{1/10}$ scaling.  Finally, the $M'$ ($4.7\mu$m) bandpass was integrated over the appropriate spectrum from the COND grid for that temperature and gravity.  In the figure, we show the results of such a calculation for $\alpha$ Cen A, the nearest sun-like star, with $L_* = 1.5L_\sun$\cite{2010MNRAS.405.1907B} assuming an age of 5 Gyr.  As expected, once irradiation becomes important the contrast of an EGP at a given separation is only weakly dependent on mass.  EGPs will be very detectable in the thermal IR in and near the HZ.

\section{NON-OPTICAL CHALLENGES}

In addition to the many optical challenges that must be overcome to image planets in the HZ, there are non-optical challenges which don't trouble current direct imaging efforts.  These are mainly due to orbital motion.  All imaging systems have a finite IWA, set by diffraction (the resolution limit) and the specifics of the coronagraph (which will be necessary).  For all but a few nearby bright stars, imaging in the HZ necessarily  implies working at small angular separation, close to the IWA.  It also implies imaging planets with comparatively short periods -- on the order of 1 year for a G star down to a few months for M stars.  The result is that a given planet may only be outside the IWA, and hence detectable, for a small fraction of the time.  Attempts to image such planets will need to account for this, by either knowing ahead of time when to look using prior knowledge (from transits, RV, or astrometry), or by repeatedly observing stars in a blind search.  For example, Brown (2004, 2005, 2010)\cite{2004ApJ...607.1003B,2005ApJ...624.1010B,2010ApJ...715..122B} has studied these issues in the context of planning for spaced-based direct imaging missions, and Shao et al (2010)\cite{2010ApJ...720..357S} analyzed the advantages prior knowledge from astrometry would give to a direct imaging search.

\begin{wrapfigure}[28]{r}{3in}
\vspace{-0.2in}
\begin{center}
\includegraphics[width=8cm]{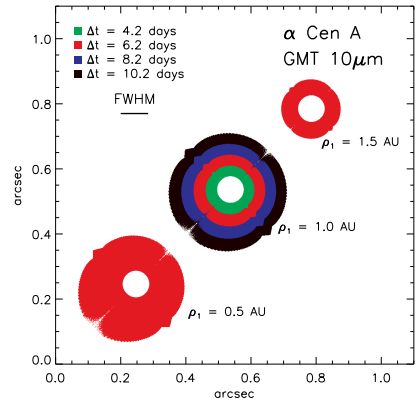}
\end{center}
\vspace{-.2in}
\caption{Possible projected orbits for hypothetical planets orbiting $\alpha$ Cen A.  The colors correspond to various elapsed times which will realistically be necessary to build long exposure times from the ground, as we consider such things as sunrise, detector efficiencies, the need for parallactic rotation, weather etc.  For this plot, the planet is assumed to be at the center of one of the circles at the beginning of the observation, and could then be anywhere within the colored region at the end of the observation.  This orbital motion results in a loss of S/N, which in turn impacts false alarm rates and search completeness.  This example, taken from Males et al (2013)\cite{2013ApJ...771...10M}, is based on an observation of $\alpha$ Cen A by the proposed TIGER instrument at GMT\cite{2012SPIE.8446E..1PH}.  Such an observation can be ``de-orbited'' to recover S/N and statistical sensitivity.
\label{fig:acena_orbs}}
\end{wrapfigure}

Orbital motion can impact direct imaging within a single observation as well, due to the motion of the planet.  As we have shown above, exoplanets are faint.  This means we can expect to need long exposure times, and predictions in the literature are often on the order of 10 hours or more \cite{2006A&A...447..397C,2010SPIE.7735E..81K}.  Such long integrations are challenging to obtain from the ground, and once we consider realistic impediments such as sunrise, weather, parallactic rotation rates, detector readouts, etc., such an observation could span several nights.  In and near the HZ orbital speed is high enough that non-negligible loss of signal-to-noise (S/N) ratio will then occur.   This is illustrated in Figure \ref{fig:acena_orbs}, which shows the trial orbits that would have to be considered during a long integration, obtained over several days, for the 25 m GMT observing $\alpha$ Cen A at $10 \mu$m.  Larger telescope diameters, short wavelengths, and lower stellar masses, all make the problem more severe.  However, Males et al (2013)\cite{2013ApJ...771...10M} showed that by ``de-orbiting'' such observations we will be able to limit the impact this loss of S/N has on the false alarm statistics and search completeness of direct imaging efforts.

\section{OBSERVATIONS OF $\alpha$ CENTAURI A WITH MagAO}

As Figure \ref{fig:ms_hz_loc} shows, EGPs in the HZ will be much easier to detect from the ground, even around G stars.  For a close enough star, the angular separation of the HZ will be wide enough that even today's AO technology with today's telescopes may be able to achieve the required contrast.  Figure \ref{fig:egp_detect} shows that irradiation will maintain the minimum brightness of an EGP in the HZ so that even very low mass gas giants will be detectable in the thermal IR.  Based on these arguments, we have begun a campaign to survey the HZs of a handful of very nearby stars using the simultaneous visible and thermal-IR imaging capabilities of the MagAO system with its VisAO and Clio2 cameras.  

MagAO is a 1 kHz, 585 actuator adaptive secondary mirror with a pyramid wavefront sensor (PWFS).  It is a near clone of the very successful LBT AO systems\cite{2010ApOpt..49G.174E,2011SPIE.8149E...1E}.  The AO system simultaneously feeds two cameras: VisAO working from 0.5-1 $\mu$m, and Clio2 working from 1-5 $\mu$m. MagAO and its two cameras were commissioned during Nov/Dec 2012 and April 2013, and hosted its first Magellan-wide science run in April of 2014.  For more information about MagAO see Morzinski et al.\cite{2014SPIE_morzinski} and Close et al.\cite{2014SPIE_close} in these proceedings.  For examples of science conducted so far with MagAO see Close et al (2013)\cite{2013ApJ...774...94C}, Follete et al (2013)\cite{2013ApJ...775L..13F}, Wu et al (2013)\cite{2013ApJ...774...45W}, Skemer et al (2014)\cite{2013arXiv1311.2085S}, Bailey et al (2014)\cite{2014ApJ...780L...4B} and Close et al (2014)\cite{2014ApJ...781L..30C}.  Of particular interest to this discussion is Males et al (2014)\cite{2014ApJ...786...32M}, where MagAO+VisAO was used to image the exoplanet $\beta$ Pictoris b in the far-red optical.  This effort made use of the simultaneous multi-wavelength imaging capability of MagAO, using a simultaneous $M'$ image of the planet to improve the significance of the optical detection.  $\beta$ Pictoris b was detected using the VisAO CCD at a contrast of $1.63\times10^{-5}$ at just 0.47''.

$\alpha$ Cen A, spectral type G2V, is the nearest sun-like star. It is somewhat more luminous than the Sun, with $L_*=1.5L_\sun$ \cite{2010MNRAS.405.1907B}, so its recent-Venus/early-Mars HZ extends from 0.9 to 2.2 AU.  Orbits in the the HZ of $\alpha$ Cen A are generally stable\cite{1997AJ....113.1445W}, and the location of the HZ is essentially unaffected by $\alpha$ Cen B\cite{2013ApJ...777..165K}.  At a distance of 1.34 pc, this projects to an angular separation of $\sim0.7$'' to $\sim1.7$'' on the sky.   $\alpha$ Cen is a triple system, of which A is the most massive.  Its close companion, $\alpha$ Cen B, is now known to host an $m\sin(i) = 1.1 M_\earth$ mass planet on a 3.3 day orbit, discovered with the RV technique\cite{2012Natur.491..207D}.  

The $\alpha$ Cen system is the target of ongoing RV monitoring campaigns.  The Anglo Australian Planet Search is complete to roughly a Saturn mass at 2 AU for $\alpha$ Cen A\cite{2011ApJ...727..102W}.  Several high cadence surveys, designed to look for lower mass terrestrial planets, are targeting the system\cite{2014arXiv1403.4809E}, though the intrinsically noisier A has detection limits which are higher than for B\cite{2011A&A...525A.140D} and it is not included in all of the high cadence projects\cite{2011A&A...534A..58P}.  These types of RV surveys hope to reach precisions which will detect large terrestrial planets in the inner HZ of $\alpha$ Cen B, but such results would only detect a $\sim$$10 M_\earth$ planet on the outer edge of the HZ of $\alpha$ Cen A.  As we have argued above, even very low mass EGPs will be detectable in the HZ, and a direct imaging search of the HZ of $\alpha$ Cen A sensitive to such planets nicely compliments the ongoing RV searches.

We observed $\alpha$ Cen A with MagAO in April 2014.  We used the $i'$ filter on VisAO, which has a central wavelength of $0.77 \mu$m, and the $M'$ filter on Clio2 at $4.7 \mu$m.  A near-focus occulting spot was used on VisAO to attenuate the light from the star.  No ND filters were required in either camera or the PWFS.  However, because of the extreme star brightness the science cameras must be operated at high frame rates.  VisAO, a frame transfer CCD, was run at 6.7 frames per second (f.p.s.) using a 512x512 field of view (FOV) with 100\% efficiency. The platescale is 7.9 mas per pixel, so this FOV is sufficient to image the full HZ.  Similarly, Clio2 integration time was set to 164 ms, but the camera readout results in only $\sim$$35\%$ efficiency. These frame rates result in huge volumes of data. The results we present here are only preliminary reductions, with a great deal of work remaining.

\begin{figure}
\begin{center}
\includegraphics[width=7.cm]{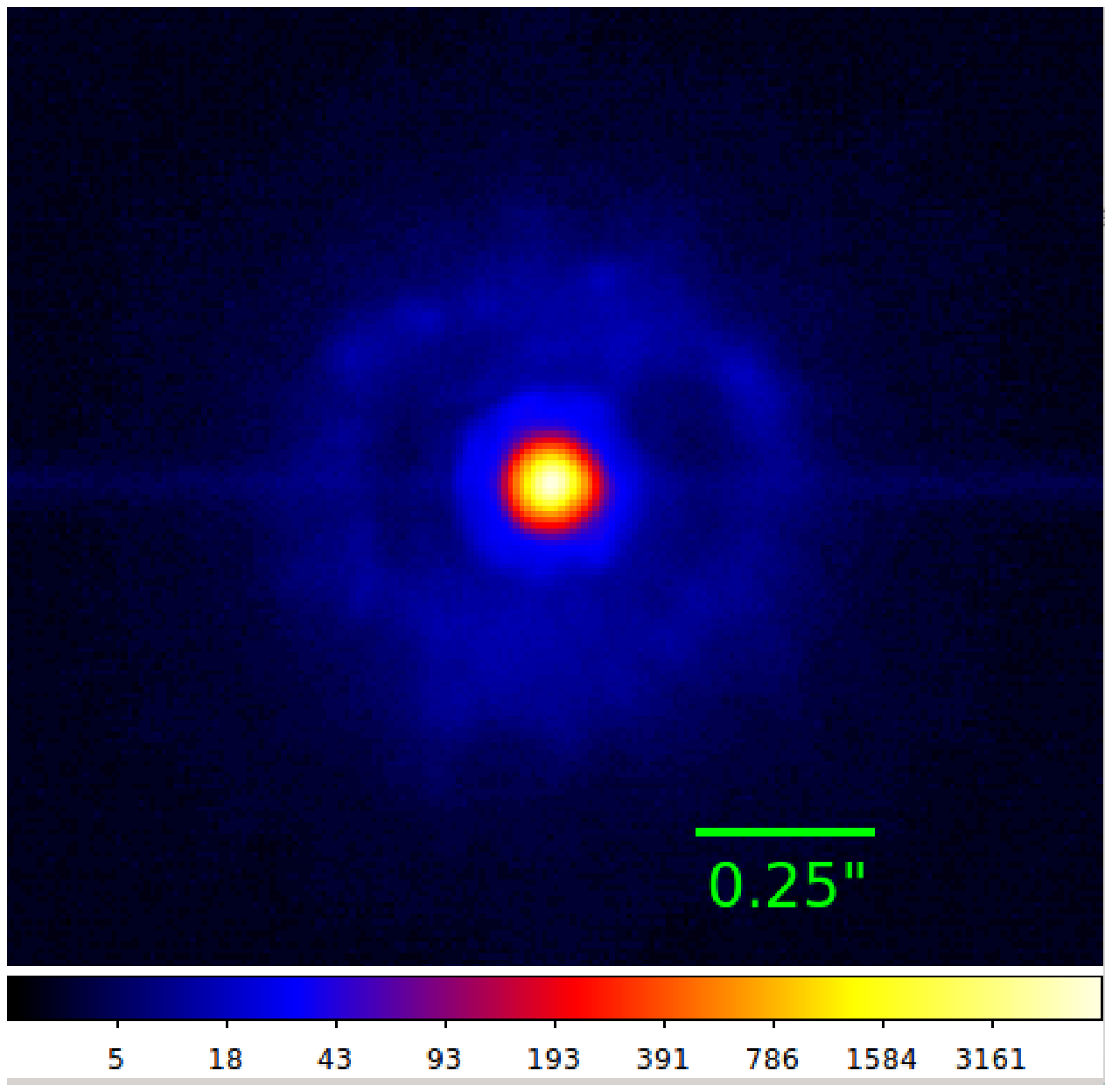}
\includegraphics[width=7.5cm]{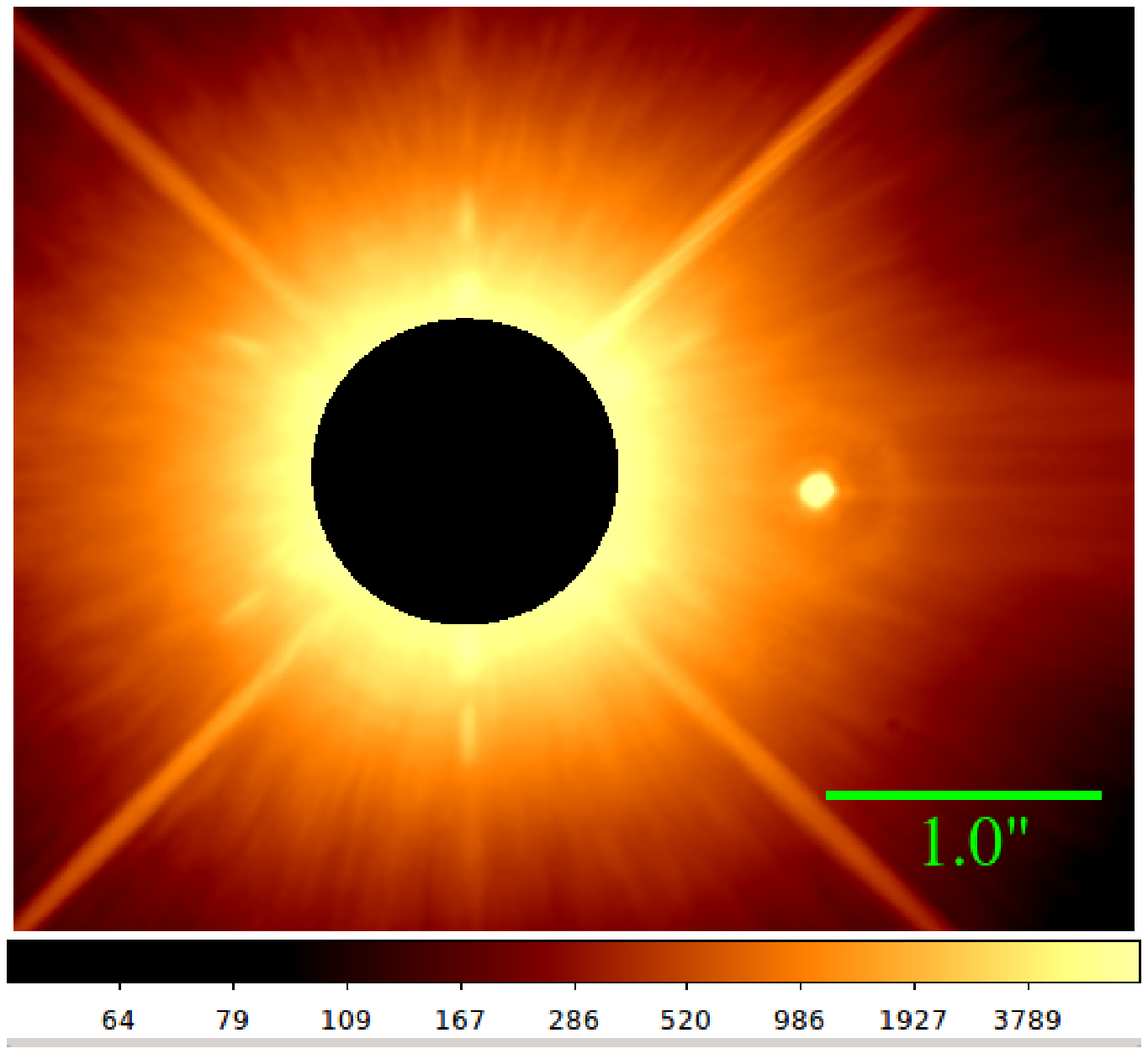}
\end{center}
\caption{MagAO+VisAO $i'$ (0.77 $\mu$m) PSF.  At left we show the unsaturated PSF taken with an ND, used for calibration.  Our preliminary Strehl ratio measurement is $\sim35\%$.  At right we show the deep pupil-aligned PSF taken behind the near-focus anti-blooming occulting mask.  This corresponds to $\sim4$ hrs of open shutter time.  The bright feature to the right is a beam splitter ghost.
\label{fig:acena_ip_psfs}}
\end{figure}

\begin{figure}
\begin{center}
\includegraphics[width=12cm]{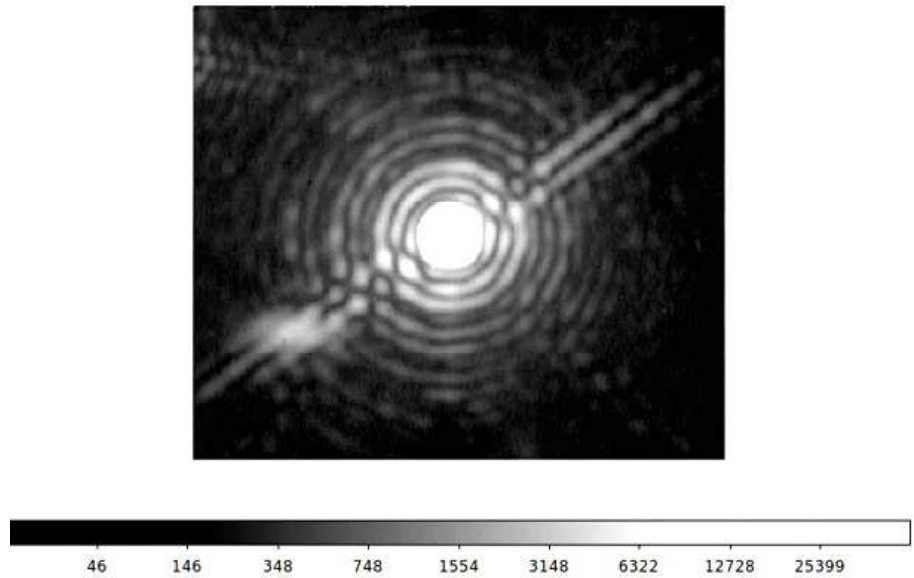}
\end{center}
\caption{MagAO+Clio2 $M'$ (4.7 $\mu$m) PSF.  This is a single 164 ms exposure, with only background subtraction applied.  The core is saturated to a radius of $0.3''$ so Strehl ratio can not be measured.  The PSF appears to be very stable at this wavelength.
\label{fig:acena_mp_psf}}
\end{figure}

In Figure \ref{fig:acena_ip_psfs} we show the point spread functions (PSFs) in $i'$.  The unsaturated PSF has a Strehl ratio of roughly $35\%$.  The occulted PSF is the combination of $\sim$$4$ hrs open shutter.  In Figure \ref{fig:acena_mp_psf} we show the Clio2 $M'$ PSF.  This is a single 164 ms frame, with only a simple background subtraction applied.  The core is saturated to $\sim$$0.3''$, so we can not measure Strehl ratio, however we note that 14 Airy rings are visible and the PSF appears to be extremely stable.

\begin{figure}
\begin{center}
\includegraphics[width=7cm]{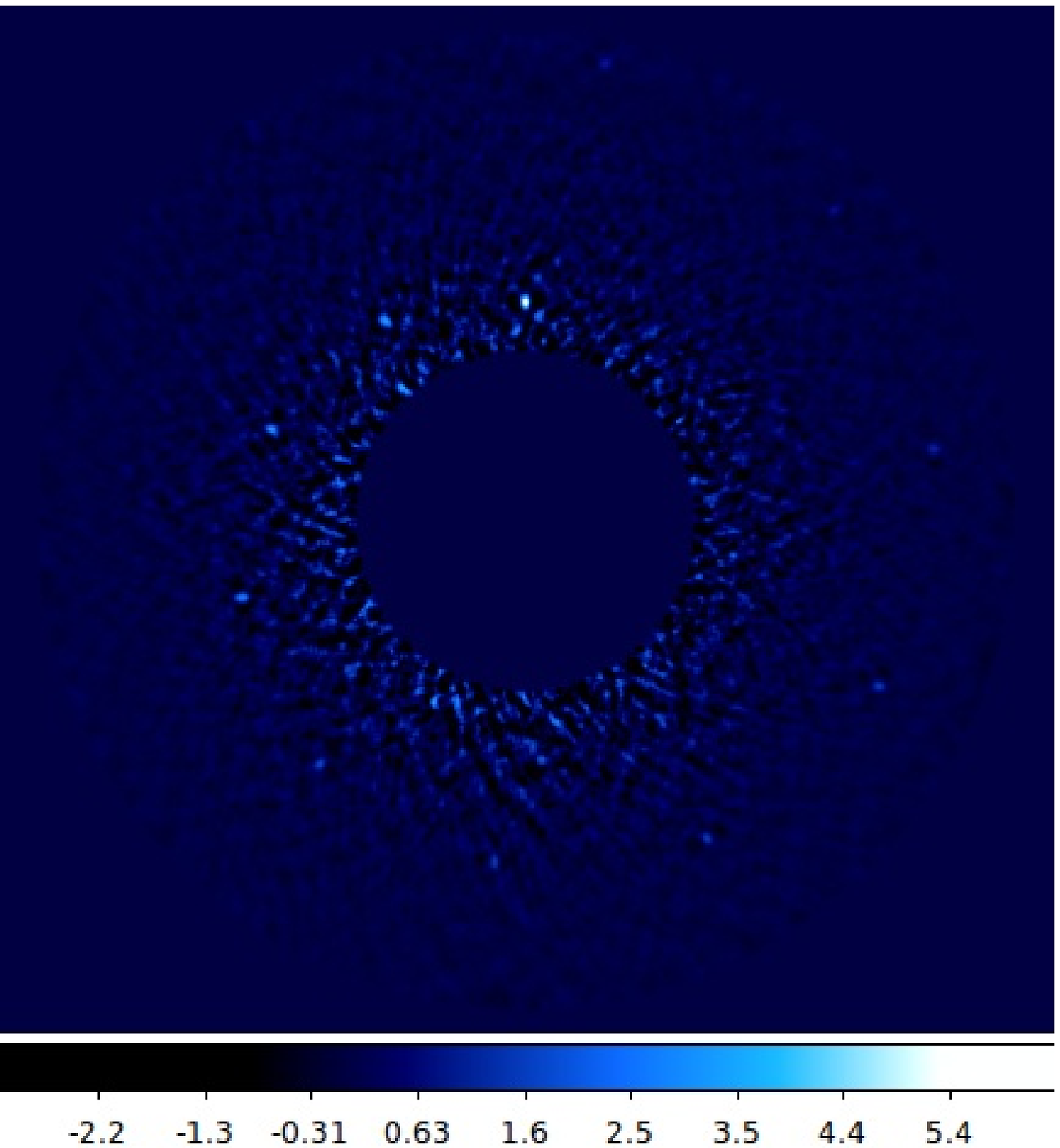}
\includegraphics[width=7cm]{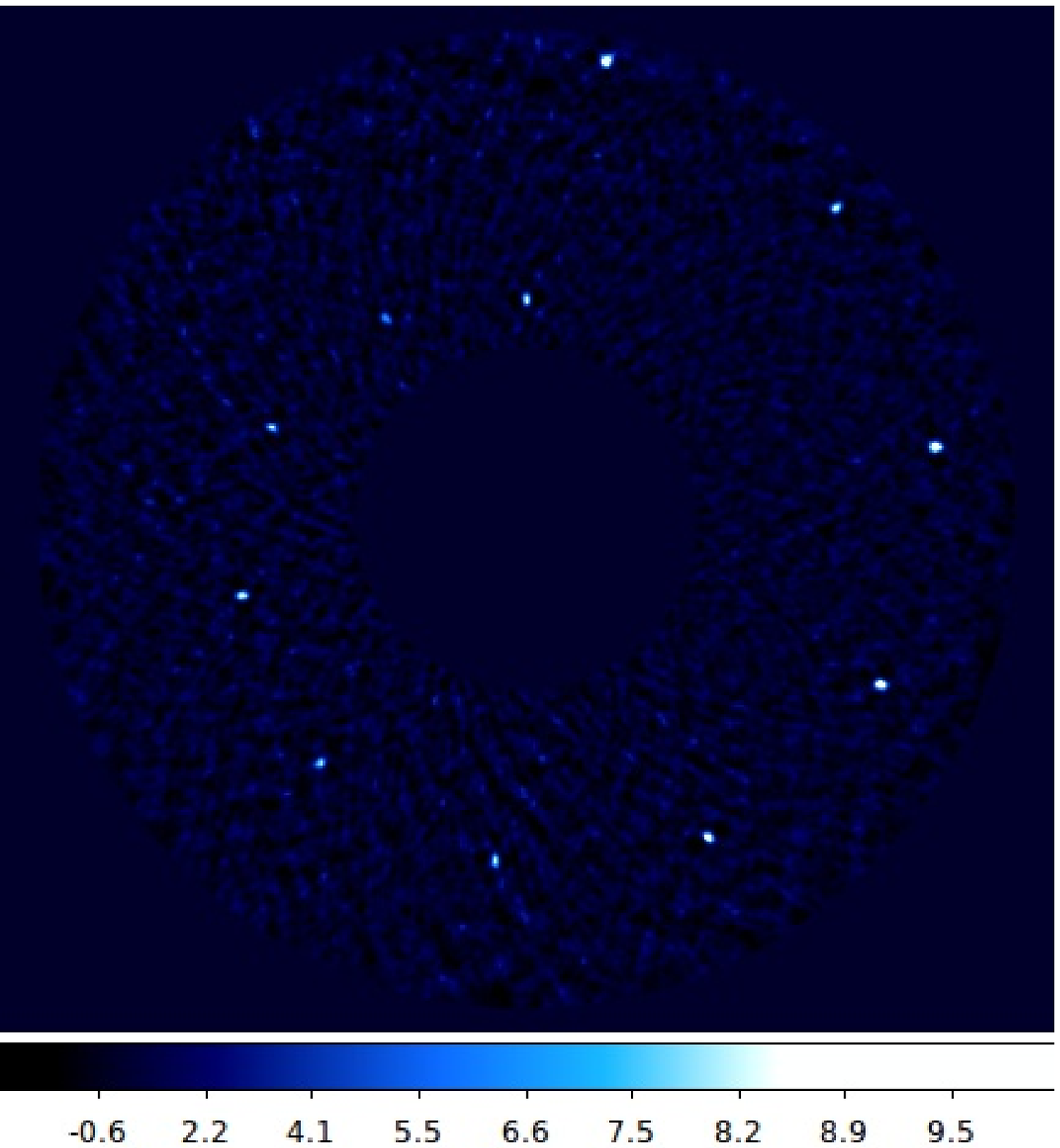}
\end{center}
\caption{Left: The habitable zone of $\alpha$ Centauri A imaged in visible light with MagAO's VisAO camera.  The data shown extends from 0.55'' to 1.58''.  Right: S/N map.  In both images the spiral of $\sim$$10\sigma$ fake planets was injected into the data to measure throughput.  The innermost planets correspond to a contrast of $\sim$$10^{-6}$, and the outermost planets correspond to a contrast of $\sim$$10^{-7}$.  These are only preliminary reductions of data taken in April, 2014, however they show that direct imaging in the HZ from the ground using AO is feasible with even today's systems.  Future extreme AO systems deployed on ELTs will further push ground based imaging into this exciting regime.    
\label{fig:acena_hz_wmagao}}
\end{figure}

In Figure \ref{fig:acena_hz_wmagao} we show our preliminary reduction of the VisAO $i'$ data.  We employed angular differential imaging and used the KLIP algorithm\cite{2012ApJ...755L..28S}.  We also show a S/N map. The spiral of fake planets was injected to measure algorithm throughput, and generally have a S/N $\sim$$10$ in this reduction.  The innermost fake planets were injected with a contrast of $\sim$$10^{-6}$, and outermost with a contrast of $\sim$$10^{-7}$ (based on a preliminary calibration).  

\section{CONCLUSIONS}

In this article we have argued that direct imaging in the HZ using ground-based AO systems will be an important part of efforts to characterize exoplanets.  Even though Earth-like planets orbiting Sun-like stars are out of reach for ground-based systems, the enormous variety of exoplanets and their host stars means that many HZ planets will be characterized using ground-based systems.  Such efforts will be complimentary to spaced-based campaigns, since the much tighter projected separations of HZs around low-mass stars will require ELT diameters to resolve.  Precursor searches with ground-based systems will also identify stars with large radius planets orbiting in or near the HZ, allowing prioritization of precious space telescope time for stars around which terrestrial planets are not ruled out by the presence of larger planets.

Imaging in the HZ is even possible using current systems for a handful of nearby stars.  Using the MagAO system and its co-mounted VisAO and Clio2 cameras we have begun a survey of nearby HZs.  Here we introduced our observations of the HZ of $\alpha$ Cen A, the nearest Sun-like star.

The MagAO ASM was developed with support from the NSF MRI program.  The MagAO PWFS was developed with help from the NSF TSIP program and the Magellan partners.  The VisAO camera and commissioning were supported with help from the NSF ATI program. The ASM was developed with support from the excellent teams at Steward Observatory Mirror Lab/CAAO (University of Arizona), Microgate (Italy), and ADS (Italy).  J.R.M. and K.M. were supported under contract with the California Institute of Technology (Caltech) funded by NASA through the Sagan Fellowship Program.  L.M.C.’s research was supported by NSF AAG and NASA Origins of Solar Systems grants. This research has made use of the Exoplanet Orbit Database and the Exoplanet Data Explorer at exoplanets.org.

\bibliographystyle{spiebib}

\bibliography{males2014}

\end{document}